\documentstyle[bib,epsf,letters]{mn}
\def\Hb{H$\beta$}
\def\C4{{\sf C4}}

\newif\ifAMStwofonts
\AMStwofontstrue

\ifoldfss
  \ifCUPmtlplainloaded \else
    \NewTextAlphabet{textbfit} {cmbxti10} {}
    \NewTextAlphabet{textbfss} {cmssbx10} {}
    \NewMathAlphabet{mathbfit} {cmbxti10} {} 
    \NewMathAlphabet{mathbfss} {cmssbx10} {} 
  \fi
  \ifAMStwofonts
    \ifCUPmtlplainloaded \else
      \NewSymbolFont{upmath} {eurm10}
      \NewSymbolFont{AMSa} {msam10}
      \NewMathSymbol{\upi}     {0}{upmath}{19}
      \NewMathSymbol{\umu}     {0}{upmath}{16}
      \NewMathSymbol{\upartial}{0}{upmath}{40}
      \NewMathSymbol{\leqslant}{3}{AMSa}{36}
      \NewMathSymbol{\geqslant}{3}{AMSa}{3E}

       \let\le=\leqslant
       \let\ge=\geqslant
    \fi
  \fi
\fi 

\ifnfssone
  \newmathalphabet{\mathit}
  \addtoversion{normal}{\mathit}{cmr}{m}{it}
  \addtoversion{bold}{\mathit}{cmr}{bx}{it}
  \newmathalphabet{\mathbfit} 
  \addtoversion{normal}{\mathbfit}{cmr}{bx}{it}
  \addtoversion{bold}{\mathbfit}{cmr}{bx}{it}
  \newmathalphabet{\mathbfss} 
  \addtoversion{normal}{\mathbfss}{cmss}{bx}{n}
  \addtoversion{bold}{\mathbfss}{cmss}{bx}{n}
  \ifAMStwofonts
    \ifCUPmtlplainloaded \else
      \UseAMStwoboldmath
      \makeatletter
      \new@mathgroup\upmath@group
      \define@mathgroup\mv@normal\upmath@group{eur}{m}{n}
      \define@mathgroup\mv@bold\upmath@group{eur}{b}{n}
      \edef\UPM{\hexnumber\upmath@group}
      \new@mathgroup\amsa@group
      \define@mathgroup\mv@normal\amsa@group{msa}{m}{n}
      \define@mathgroup\mv@bold\amsa@group{msa}{m}{n}
      \edef\AMSa{\hexnumber\amsa@group}
      \makeatother
      \mathchardef\upi="0\UPM19
      \mathchardef\umu="0\UPM16
      \mathchardef\upartial="0\UPM40
      \mathchardef\leqslant="3\AMSa36
      \mathchardef\geqslant="3\AMSa3E

       \let\le=\leqslant
       \let\ge=\geqslant
    \fi
  \fi
\fi 

\ifnfsstwo
  \DeclareMathAlphabet{\mathbfit}{OT1}{cmr}{bx}{it}
  \SetMathAlphabet\mathbfit{bold}{OT1}{cmr}{bx}{it}
  \DeclareMathAlphabet{\mathbfss}{OT1}{cmss}{bx}{n}
  \SetMathAlphabet\mathbfss{bold}{OT1}{cmss}{bx}{n}
  \ifAMStwofonts
    \ifCUPmtlplainloaded \else
      \DeclareSymbolFont{UPM}{U}{eur}{m}{n}
      \SetSymbolFont{UPM}{bold}{U}{eur}{b}{n}
      \DeclareSymbolFont{AMSa}{U}{msa}{m}{n}
      \DeclareMathSymbol{\upi}{0}{UPM}{"19}
      \DeclareMathSymbol{\umu}{0}{UPM}{"16}
      \DeclareMathSymbol{\upartial}{0}{UPM}{"40}
      \DeclareMathSymbol{\leqslant}{3}{AMSa}{"36}
      \DeclareMathSymbol{\geqslant}{3}{AMSa}{"3E}

       \let\le=\leqslant
       \let\ge=\geqslant
    \fi
  \fi
\fi 

\ifCUPmtlplainloaded \else
  \ifAMStwofonts \else 
    \def\upi{\pi}
    \def\umu{\mu}
    \def\upartial{\partial}
  \fi
\fi

\title{The Shapes and Ages of Elliptical Galaxies}
\author[Roelof S. de Jong and Roger L. Davies]
       {Roelof S. de Jong and Roger L. Davies\\
        University of Durham, Department of Physics, South Road, 
        Durham DH1 3LE, England}
\date{Accepted 
      Received 
      in original form }

\pagerange{\pageref{firstpage}--\pageref{lastpage}}
\pubyear{1996}

\begin{document}

\maketitle

\label{firstpage}

\begin{abstract}
In this paper we investigate the relation between the detailed isophotal
shape of elliptical galaxies and the strength of the \Hb\ absorption in
their spectra. We find that disky galaxies have higher \Hb\ indices.
Stellar population synthesis models show that the \Hb\ line is a good age
indicator, hence disky galaxies tend to have younger mean ages than boxy
galaxies.  We show that the observed trend can be brought about by a
contaminating young population, which we associate with the disky
component. This population need only account for a small fraction of the
total mass, for example if a contaminating population of age of 2\,Gyrs is
superimposed on an old (12\,Gyr) elliptical galaxy, then the observed
trend can be explained if it contributes only 10\% to the total mass. The
size of this effect is consistent with the estimates of disk-to-total
light ratios from surface photometry.

\end{abstract}

\begin{keywords}
galaxies: elliptical and lenticular --
galaxies: evolution --
galaxies: stellar content --
galaxies: structure 
\end{keywords}

\section{Introduction}

Recently a controversy has arisen over the ages of elliptical galaxies. 
The conventional view is that elliptical galaxies are old and coeval
systems, formed by an initial starburst about 15\,Gyrs ago.  New
population syntheses (Worthey~\cite{Wor94}) combined with new
observations of linestrengths in a large sample of elliptical galaxies
(Gonz\'alez~\cite{GonPhD}), have brought this view into question (Faber
et al.~\cite{Fab95}; Worthey et al.~\cite{Wor95}; for review see
Davies~\cite{Dav96}).  These studies show that over a small range in
metallicity, luminous elliptical galaxies span a wide range of \Hb\
absorption strengths implying ages from greater than 12 Gyrs to as young
as 3 Gyrs.  Many elliptical galaxies have isophote shapes that deviate
from pure ellipses at the 1\% level in the sense of being either
``disky'' or ``boxy'' (Bender et al.~\cite{Ben89}).  Here we investigate
the hypothesis that the apparent young ages are brought about by the
superposition of a young population (associated with the disky
component) on an old elliptical population. 

Determinations of the ages and metallicities of elliptical galaxies
suffer from a degeneracy, both increased age and increased metallicity
tend to redden the spectral energy distribution of a single stellar
population and strengthen the metal absorption lines
(Renzini~\cite{Ren86}; Buzzoni et al.~\cite{Buz92};
Worthey~\cite{Wor94}). Recently, spectral indices have been identified
that break this degeneracy.  The stellar population synthesis models of
Worthey~(\cite{Wor94}) indicate the \Hb\ and [MgFe] indices can be used
in combination to determine age and metallicity separately.
Gonz\'alez~(\cite{GonPhD}) collected longslit spectra of major and minor
axis of a sample of about 40 elliptical galaxies. His observations
indicate a substantial age spread among elliptical galaxies, as the \Hb\
index shows a large range of values over a small range in metallicity.
This conclusion is difficult to reconcile with the hypothesis that
elliptical galaxies have a single epoch of formation and are old, as has
been inferred from the small scatter in their colour--magnitude,
Fundamental Plane and $Mg_2, \sigma$ relations (Bower
et al.~\cite{Bow92a},~\cite{Bow92b}; Renzini~\cite{Ren95}).
However in cosmological simulations where galaxies form through a
hierarchy of mergers (Cole et al~\cite{Cole94}; Kauffmann~\cite{Kau95};
Baugh, Cole \& Frenk~\cite{Bau96}), the stars that currently dominate
the luminosity of giant galaxies could have a range of ages (if we
associate star formation with those mergers that are gaseous). In such
models, luminous elliptical galaxies form late (at redshifts $<$\,1)
suggesting that at least the youngest populations, formed in the most
recent merger, would be $<$\,5\,Gyrs old. If elliptical galaxies consist of
populations of different ages, the linestrengths will only reflect the
{\it luminosity weighted\,} average age, not the age of formation or even
the age of the last starburst.  As young populations tend to be much more
luminous, a small young population can dramatically change the strength
of the lines (for detailed modeling see Charlot \& Silk~\cite{ChaSil94};
Bressan, Chiosi \& Tantalo~\cite{Bre96}). If there are several stellar
population components in a galaxy one has to remain alert to the
possibility that the \Hb\ and [MgFe] lines might not arise from the same
stellar populations.

Elliptical galaxies have almost perfectly elliptical isophotes however
small deviations from ellipses have been measured repeatably (Peletier et
al.~\cite{Pel90}, Goudfrooij et al.~\cite{Gou94}).  Higher order Fourier
terms have been used to parametrise these deviations.  Unless isophotes
are dramatically perturbed by the presence of dust, the higher order
terms are small, only the fourth order cosine (cos(4$\theta$))
term showing significant deviations from zero.  The maximum deviation in
the cos(4$\theta$) term as function of radius is taken as the
characteristic value for the system, denoted by \C4. Galaxies with disky
isophote distortions have positive \C4\ values and those with boxy
isophote distortions have negative values of \C4.

Projection effects make the general interpretation of the \C4\ parameter
non-trivial (Kochanek \& Rybicki~\cite{KocRyb96}; Gerhard \&
Binney~\cite{GerBin96}).  The \C4\ interpretation of highly inclined
systems is less ambiguous, positive \C4's clearly result from disks
embedded in the galaxy body.  Even so, the \C4 values cannot be
translated directly into a disk-to-bulge (D/B) ratio.  \C4\ only relates
to the maximum deviation from ellipticity at a certain radius and the
relation between \C4 and D/B ratio depends critically on the relative
surface brightness and scalelengths of bulge and disk.  We will use \C4\
values only as an indicator of disk presence, not as an direct indicator
for B/D ratios.  

Here we explore how the diagnostic of youth, {\Hb~strength}, is correlated
with the deviation of the isophotes from perfect ellipses. In section two
we present the correlation and in section three we explore how such
a trend can be accounted for using Worthey's population models. In section
4 we consider the wider implications.

\section{\Hb\ index versus isophotal shape}

To investigate whether there is a relation between isophotal shape
and the \Hb\ index for elliptical galaxies, a large number of
measurements were collected from the literature. 


The \C4 isophotal shape parameters were collected from Bender et
al.~(\cite{Ben89}), Peletier et al.~(\cite{Pel90}) and Goudfrooij et
al.~(\cite{Gou94}). Two different methods to evaluate \C4 were used by
these authors, and the difference produces a small scaling between the
measured values.  This was removed using a linear relation for the
galaxies in common between the different authors (see also
Peletier~\cite{PelPhD}). The maximum rms error between the measured \C4
parameters of any two papers was 0.003, which we will assume to be the
typical uncertainty in the \C4 parameter.  The average of all available
\C4 measurements ($<$\C4$>$) was used as best estimate of the \C4
parameter of a galaxy. 

The measurements of the equivalent width of the \Hb\ absorption line
were mainly obtained from Gonz\'alez~(\cite{GonPhD}), with a few
additional galaxies from Davies, Sadler \& Peletier~(\cite{Dav93}). 
Gonz\'alez simulated aperture spectroscopy of his galaxies using
longslit spectra.  Azimuthal integration was simulated by taking
luminosity and radially weighted averages of the linestrengths measured
along the major and minor axis.  We applied a similar procedure for the
Davies et al.\ data, calculating linestrengths within the 1/8 and the
1/2 effective radius ($R_{\rm e}$) of the galaxies.  Gonz\'alez
corrected the \Hb\ absorption linestrengths for infilling due to
emission by a relation between the strength of the \Hb\ line emission
and that of the the [O{\sc III}] line at 5007\,\AA, namely $\Delta$\Hb =
0.7\,[OIII].  We have adopted these corrected values.  The validity of
this correction has been questioned by Carrasco et al.~(\cite{Car95})
who compared the change in \Hb\ with that in O[III] from one side of the
nucleus to an equal radius on the opposite side in a sample of 26
galaxies.  Assuming that the intrinsic \Hb\ absorption is symmetric about
the nucleus they should have recovered the Gonz\'alez relationship,
however they did not, rather they found no correlation between the
changes for galaxies with [O{\sc III}]\,$<$\,1.5\,\AA.  The emission
correction is rather uncertain as the range of O[III]/\Hb\ emission
ratios in LINERS and H{\sc II} regions is large (Baldwin, Phillips,
Terlevich~\cite{Bal81}; Ho, Filippenko \& Sargent~\cite{Ho93}).  We use
the corrected values but note that the uncorrected (lower) values of \Hb\
result in age estimates that are considerably older than those discussed
by Gonz\'alez and collaborators and therefore establishing the correct
\Hb\ emission correction is a fundamental step.  In general the
measurements of Gonz\'alez and Davies et al.\ agree to within the stated
errors, with a typical uncertainty of 0.18\,\AA\ rms. 

\begin{figure}
 \mbox{\epsfxsize=8.6cm\epsfbox[30 165 570 550]{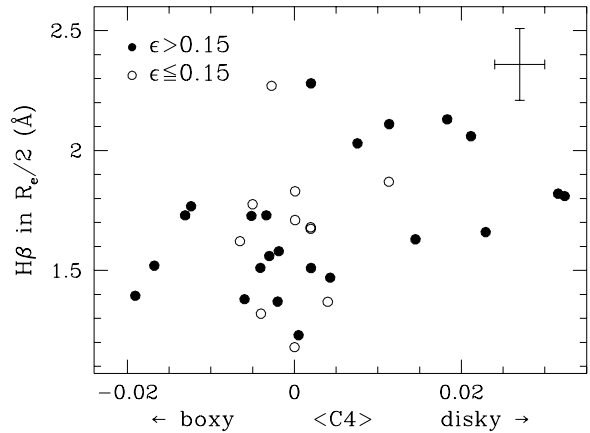}}
\caption{
 The average \C4 index, determined as described in text, versus the
\Hb\ equivalent width within $R_{\rm e}/2$. The open symbols indicate
galaxies with ellipticity ($\epsilon$) less than 0.15, solid symbols
have ellipticity more than 0.15, for which it is easier to determine
\C4. Typical error estimates are indicated in top-right corner.
}
 \label{c4hb2}
 \end{figure}

In Fig.~\ref{c4hb2} we plot the $<$\C4$>$ parameter against the \Hb\
equivalent width within $R_{\rm e}/2$ for all the galaxies for which we
have data.  For galaxies with ellipticities of $\epsilon\le0.15$ it is
hard to determine \C4, these are indicated by open circles.  A trend is
apparent in Fig.~\ref{c4hb2}, for galaxies with $\epsilon > 0.15$ the
Spearman rank correlation coefficient is 0.48 indicating that the
probability of finding such a correlation by chance is less than 2.5\%. 
All boxy galaxies have \Hb\ indices less than 1.8\,\AA, while the disky
galaxies have often much higher indices, with no values below 1.6\,\AA. 
Similar trends are found when $R_{\rm e}/8$ is used instead of $R_{\rm
e}/2$ or when the \Hb\ linestrengths are not corrected for emission.

\section{Models}

Can the $<$\C4$>$ versus \Hb\ trend be understood in terms of simple
stellar synthesis models by combining an old and a young
population? If the high \Hb\ indices in disky galaxies are caused by a
young disk component superposed on the old bulge component, we have to
determine the typical mass and/or luminosity fraction of this disk
within $R_{\rm e}/2$ or $R_{\rm e}/8$.  A number of disk/bulge
decomposition methods can be found in the literature (for review see
Capaccioli \& Caon~\cite{CapCao92}).  We will use the 2D bulge/disk
decompositions of Simien \& Michard (\cite{SimMic90}) and of Scorza \&
Bender (\cite{ScoBen95}) to determine the typical range in Disk-to-Total
(D/T) luminosity fraction of disky ellipticals in the region where the
\Hb\ indices have been measured. 

Simien \& Michard find D/T ratios of 0.14 for two elliptical galaxies. 
Scorza \& Bender find D/B ratios from 0.03 to 0.28 for their sample of
disky ellipticals, which translates in D/T ratios of about 0.03 to 0.22. 
They were able to calculate {\it local} D/B ratios and concluded that
these values were consistent with the values they derived from the
independent analysis of line profile shapes determined from long slit
spectra.  They found typical local D/T values within $R_{\rm e}/2$ of
0.1 to 0.4 for disky ellipticals.  We will use this range of values to
form models of old and young populations using the models of Worthey
(\cite{Wor94}).  Because younger populations are more luminous than
older ones we use the $M/L$ ratios of the models to link the observed
luminosity ratios to the mass ratios of the populations.  Similarly we
cannot simply take the average of the equivalent \Hb\ (or other)
linewidths of the old and the young population to calculate the
linestrengths of the combined population.  Ideally we should add the
spectra of the two populations scaled by the appropriate mass fractions,
however as Worthey's models do not synthesise spectra but give
magnitudes, colours and line indices, we have adopted a simpler
approach.  Before averaging two populations, we weight the linestrengths
by the local continuum, which we approximate by using the $M/L_V$ ratios
of the young and old populations.  Starting with a 12\,Gyr old
population, we add a young population of between 0 and 20\% in mass to
calculate the resulting D/T luminosity ratios and linestrengths. 

 \begin{figure} 
 \mbox{\epsfxsize=8.6cm\epsfbox[30 165 570 550]{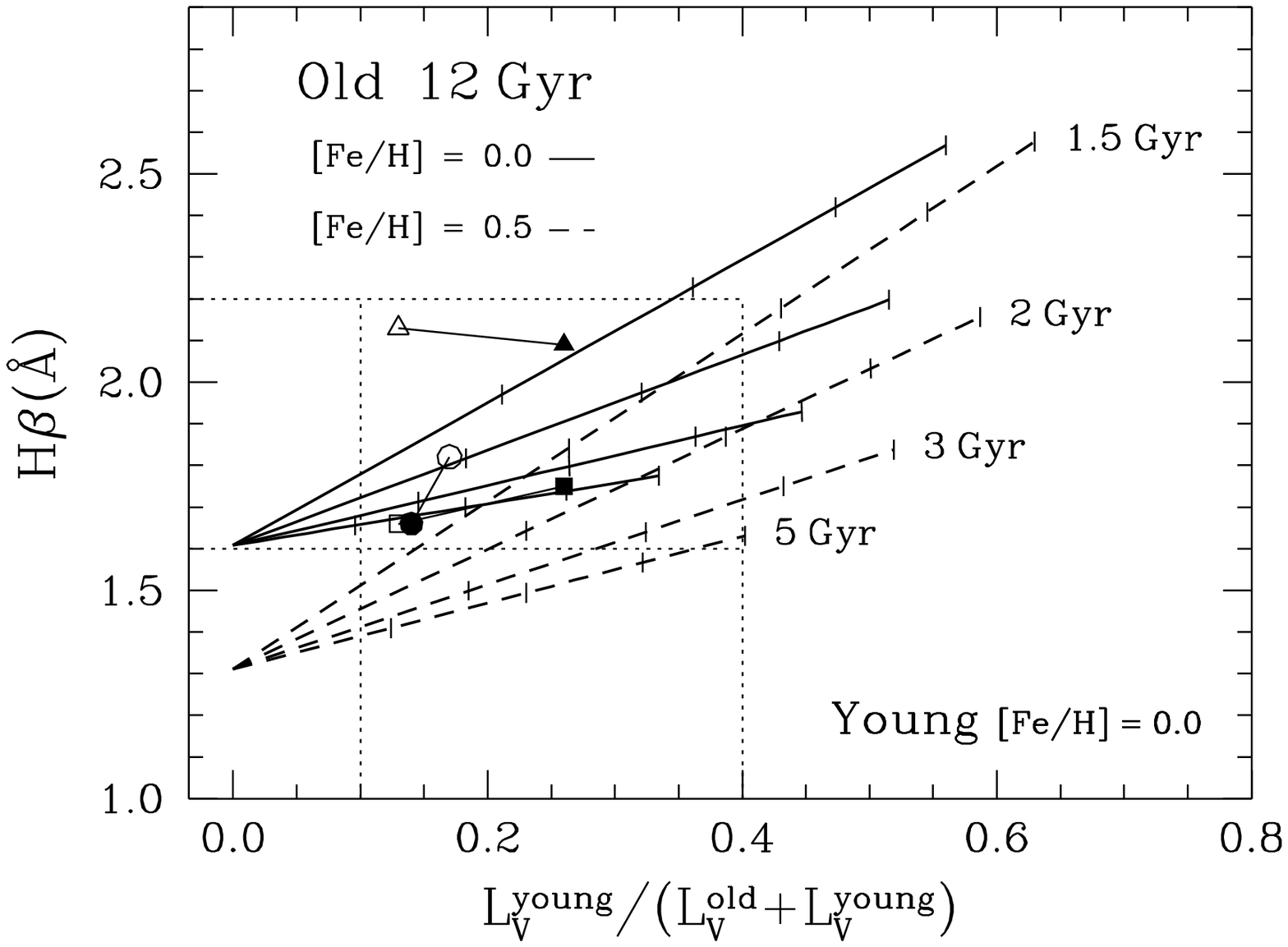}}
 \mbox{\epsfxsize=8.6cm\epsfbox[30 165 570 550]{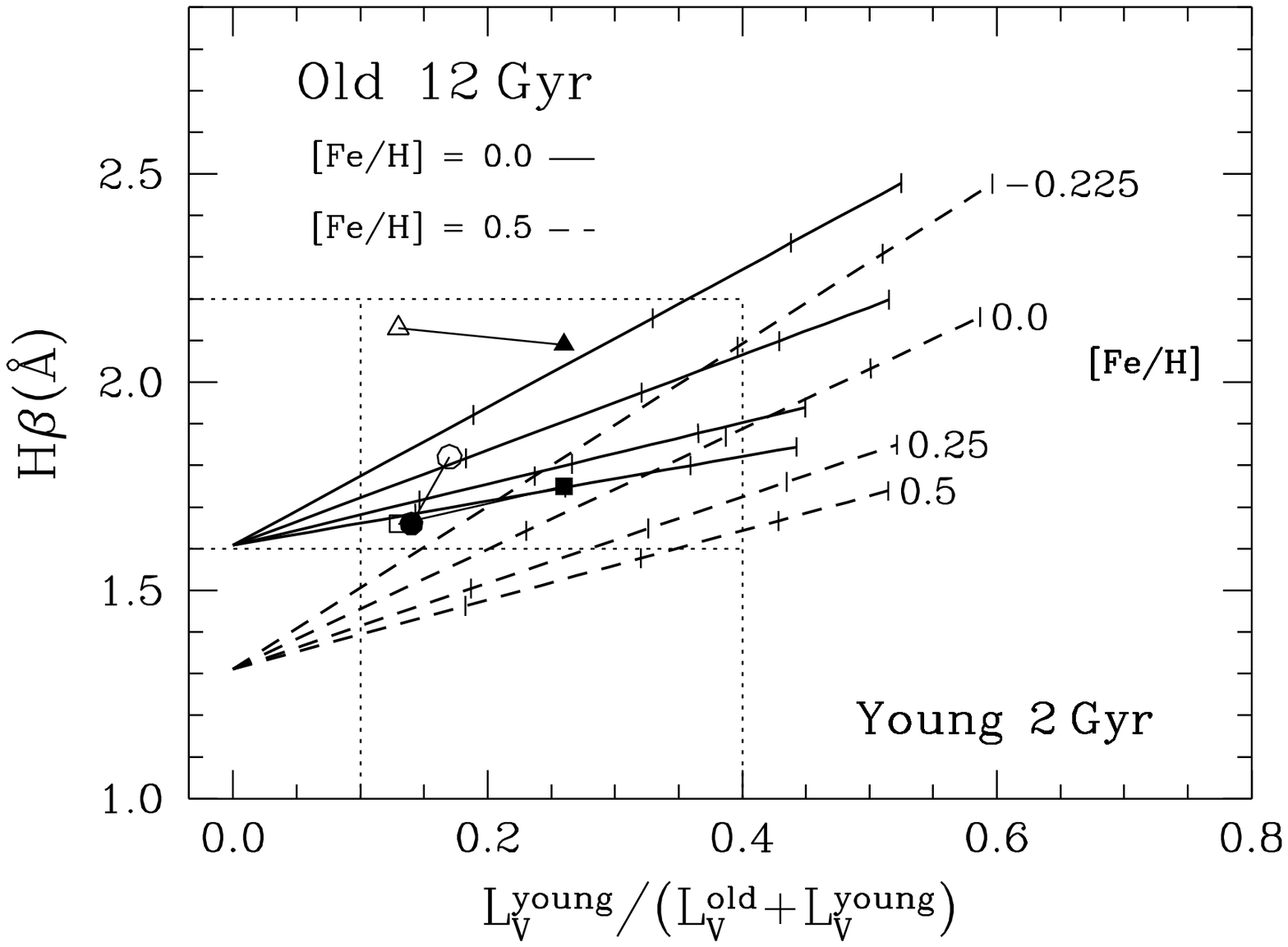}}
 \caption{ Model calculations of the combined effects of old and young
populations, showing the young-to-total luminosity ratio versus \Hb\
equivalent width.  All old populations are 12\,Gyr, with either
metallicity [Fe/H]=0.0 or 0.5 as indicated.  In the top panel the young
population is of solar metallicity, with ages 1.5, 2, 3 or 5\,Gyr from
top to bottom.  In the bottom panel all young populations have an age of
2\,Gyr, but the metallicity varies with [Fe/H]=-0.225, 0.0, 0.25 and 0.5
from top to bottom.  Each tickmark along the model lines corresponds to
a 5\% increase in the young-to-total mass of the stellar populations. 
The dotted lines indicate the observed range in \Hb\ linestrength and
luminosity ratios in disky ellipticals.  Data for three galaxies
(NGC\,821 circles, NGC\,3377 triangles and NGC\,4268 squares) are
plotted, the solid symbols indicate the $R_{\rm e}/8$ values, the open
symbols the $R_{\rm e}/2$ values (see text). 
}
 \label{modrat}
 \end{figure}

Figure~\ref{modrat} shows the results of such calculations.  Starting
with a 12\,Gyr old population of either solar or about 3 times solar
metallicity, we add, in the top figure, solar metallicity populations of
1.5, 2, 3 or 5\,Gyr and, in the bottom figure, 2\,Gyr populations with
[Fe/H]\,=\,-0.225, 0, 0.25 or 0.5.  The mass fraction of young-to-total
population was varied from 0 to 20\%. The young populations have a strong
effect on the luminosity fraction and the indices, even for small mass
fractions.  Different age and metallicity combinations for old and young
populations illustrate the same basic result. A 3\,Gyr old population can
easily increase the \Hb\ index by 0.4--0.8\,\AA, even with mass fractions
of 20\% or less.  Making the old population even older than 12\,Gyr will
shift the tracks down slightly so that a smaller young mass fraction is
needed to produce the same effect in luminosity and \Hb.  \Hb\ is a good
age indicator, with less sensitivity to metallicity than most other
lines, but a lower metallicity young population still has considerably
more effect on the \Hb\ equivalent width than a high metallicity young
population.

Let us now compare the observed range in \Hb\ indices and D/T ratios
with these models.  All boxy galaxies have \Hb\ indices less than
1.8\,\AA, consistent with the single burst 12\,Gyr old 
populations of solar or super-solar metallicity.  The disky galaxies
have \Hb\ indices of 1.6--2.2\,\AA\ within $R_{\rm e}/2$ (and similar
values within $R_{\rm e}/8$), while the D/T ratios are between 0.1 and
0.4 in these simulated apertures.  These ranges have been indicated by
the dotted lines in Fig.~\ref{modrat}.  Clearly the full range in \Hb\
can be reached within the permitted range in D/T light ratios.  In the
most extreme case of a (12\,Gyr old plus a 1.5\,Gyr young) population, a 10\%
disk mass contamination is sufficient to cover the full \Hb\ range. 

For three galaxies (NGC\,821, NGC\,3377 and NGC\,4697) we have
calculated the exact position in Fig.~\ref{modrat}, using the D/B
decompositions of Scorza \& Bender~(\cite{ScoBen95}).  The procedure to
calculate D/B ratios within $R_{\rm e}/2$ and $R_{\rm e}/8$ was
identical that used by Gonz\'alez to calculate line indices in simulated
apertures.  D/B profiles along major and minor axis were
extracted from the original D/B maps produced by Scorza \& Bender. These
were luminosity and radially weighted to calculate average D/T values
within $R_{\rm e}/2$ and $R_{\rm e}/8$.  NGC\,821 and NGC\,4697 lie
in the region covered by the models, only the $R_{\rm e}/2$ point of
NGC\,3377 can not be explained by the models. This galaxy has a large
(and uncertain) emission line correction of 0.35\,\AA, which might
explain the discrepancy. A sub-solar metallicity in the main body of this
galaxy would also remove the discrepancy.

 \begin{figure}
 \mbox{\epsfxsize=8.7cm\epsfbox[30 175 570 673]{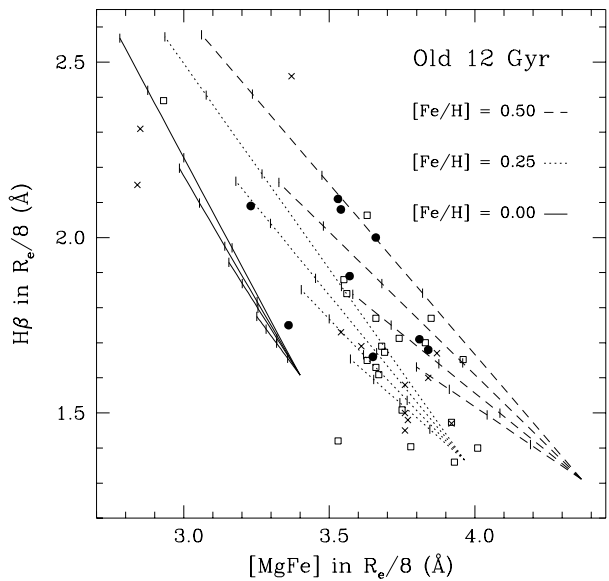}}
 \caption{
 The  \Hb\ versus [MgFe] equivalent width for all galaxies, solid circles
the galaxies with $<$\C4$>\,\ge\,0.005$, open squares $<$\C4$>\,<\,0.005$ and
crosses galaxies with no measured \C4.  The lines show models, in
which we combine 12\,Gyr old populations of the indicated metallicities
with young populations of Solar metallicity.  To each old population we
have added, from top to bottom, a 1.5, 2, 3, or 5\,Gyr young population. 
Each tickmark along the model lines is 5\% increase in the
young-to-total mass of the combined populations. 
 }
 \label{lineline}
 \end{figure}

We have shown that the modeled \Hb\ indices are consistent with the
young disk interpretation and now explore consistency with the metal line
indices.  Gonz\'alez~(\cite{GonPhD}; see also Faber et al.~\cite{Fab95})
uses a combination of the Mg{\it b} and the Fe\,5270+Fe\,5335 indices to
formulate a [MgFe] index, empirically estimated to be a good metallicity
indicator.  Figure~\ref{lineline} shows the data for Gonz\'alez galaxies
with the above models superimposed.  Consistent models for both a good
age indicator (\Hb) and a metallicity indicator ([MgFe]) can be produced
by combining an old metal-rich population with younger populations. 
Different combinations of age and metallicity and of old and young
populations can cover the distribution of points without conflicting with the
observed D/B decomposition results. 

Due to the discreteness in age and metallicity of the populations
calculated by Worthey (\cite{Wor94}) we have opted here for simple
combinations of young and old populations, however, continuous low level
star formation or a series of small starbursts will produce similar
effects in disk-to-total luminosity and integrated line indices.  The
point is that the old galaxy with young disk interpretation is
consistent with the observations.  One should be careful to interpret
the line indices observations, which are heavily biased to the central
regions of the galaxies, as indicative for the stellar population of the
whole galaxy.

\section{Discussion}

We have shown that the central \Hb\ indices of elliptical galaxies are
correlated with their \C4 parameters measuring isophotal shape.  This
suggests that the young ages found in some elliptical galaxies are
associated with the disky components in these galaxies.  Simple
modelling, adding a small young population to an old population is
capable of reproducing the change in the \Hb\ and metallicity indices by
the required amount.  These models are also consistent with
the D/T light ratios determined from the limited surface photometry
available.

Given that the disky population of ellipticals is younger than the main
body of the galaxy we are prompted to speculate on its origin.  If it
arises via mass lost from the host population that has settled into a
disk by dissipation, we expect the gas to have high metallicity. 
However if the gas was accreted either from a low mass satellite galaxy
or a reservoir of cold intergalactic gas we would expect a metallicity
lower than that of the host.  Disks formed from gas that has condensed
out of cooling flows will also be metal poor, but will form over a
extended period of time.  Therefore if we were able to determine the
metallicity of the disk (or gas) in a sample of such galaxies we would
learn about the origin of these features.  In either case not all
elliptical galaxies are dormant star-piles.  The possible range in disk
age and metallicity is strongly dependent on the disk origin. 

We conclude that further investigation of the ages of (disky) elliptical
galaxies is required, using spatially resolved age indicators.  Two
possibilities are the comparison of the spatial distribution of line
indices using long slit or integral field spectroscopy of galaxies and
the colours of disks/bulges. 

\section*{Acknowledgments}

We thank Cecilia Scorza for providing us the D/B ratio images of three
galaxies. We thank Guy Worthey for making his models available to us.

\bsp

\label{lastpage}

\end{document}